# Optimal Fusion Transformations for Linear Optical Cluster State Generation

**D. B. Uskov**[1,2*], **P. Lougovski**[3], **P. M. Alsing**[4], **M. L. Fanto**[4], **L. Kaplan**[2], **and A. Matthew Smith**[3,4]

[1]Department of Mathematics and Natural Sciences, Brescia University, Owensboro, Kentucky 42301, USA
[2]Department of Physics and Engineering Physics, Tulane University, New Orleans, Louisiana 70118, USA
[3]Quantum Information Science Group, Oak Ridge National Laboratory, Oak Ridge, Tennessee 37831, USA
[4]Air Force Research Laboratory, Information Directorate, Rome, New York 13440, USA

[*]E-mail: dmitry.uskov@brescia.edu

**Abstract.** We analyze the generation of linear optical cluster states (LOCS) via addition of one and two qubits. Existing approaches employ the stochastic linear optical two-qubit CZ gate with success rate of 1/9 per fusion operation. The question of optimality of the CZ gate with respect to LOCS generation remains open. We report that there are alternative schemes to the CZ gate that are exponentially more efficient and show that sequential LOCS growth is globally optimal. We find that the optimal cluster growth operation is a state transformation on a subspace of the full Hilbert space. We show that the maximal success rate of fusing $n$ photonic qubits or $m$ Bell pairs is $(1/2)^{n-1}$ and $(1/4)^{m-1}$ respectively and give an explicit optical design.

Cluster states [1] of photonic qubits are a fundamental resource for quantum information processing [2]. Constructing these states presents a major experimental and theoretical challenge because known physical implementations of optical multi-qubit transformations are intrinsically probabilistic [3]. To evaluate the efficiency of such transformations the success probability of a desired measurement outcome is routinely used. Thus, finding entangling photonic transformations that achieve the best possible success probability is of critical importance for progress in the field. Unfortunately, the problem of designing an optimal linear optical transformation is at least #P-complete [4]. In this Letter we construct an analytical solution to this problem for the case of linear optical cluster state generation via sequential addition of one- and two-qubit states.

We define a linear optical device operating on $M$ modes as a unitary transformation $U$ of photon creation operators from the input modes $a_i^\dagger$ to the output modes $\tilde{a}_j^\dagger$

$$a_i^\dagger \to \sum_{j=1}^{M} U_{ij} \tilde{a}_j^\dagger . \quad (1)$$

The mode transformation matrix $U$ induces a state transformation $\Omega$ which generally is a high-dimensional unitary representation of $U$ [5]. For a multi-photon product input state $|\Psi^{(in)}\rangle = \prod_{i=1}^{M} a_i^{\dagger n_i}/\sqrt{n_i!}|0\rangle$ the action of $\Omega$ will result in the output state

$$|\Psi^{(out)}\rangle = \Omega|\Psi^{(in)}\rangle = \prod_{i=1}^{M}\left(\sum_{j=1,M} U_{ij}\tilde{a}_j^\dagger\right)^{n_i} / \sqrt{n_i!}|0\rangle .$$

Next, we define logical qubit states $|\uparrow\rangle$ and $|\downarrow\rangle$ using two-mode Fock states $|\uparrow\rangle = |1,0\rangle$ and $|\downarrow\rangle = |0,1\rangle$ which may conveniently be implemented as horizontal $|H\rangle$ and vertical $|V\rangle$ photon polarizations in a single spatial mode. For an arbitrary $N$-qubit input state, the output state produced by the action of $\Omega$ will not lie in the computational space $span\left[\{|H\rangle,|V\rangle\}^{\otimes N}\right]$ but rather span a larger space of $N$ photons distributed over $M = 2N$ modes, i.e.,

$$|\Psi^{(out)}\rangle = \alpha|\Psi_I^{(out)}\rangle + \beta|\Psi_{II}^{(out)}\rangle, \quad |\alpha|^2 + |\beta|^2 = 1, \quad (2)$$

where $|\Psi_I^{(out)}\rangle$ and $|\Psi_{II}^{(out)}\rangle$ belong to the computational space and its complement, respectively. The goal is to find a mode transformation matrix $U$ such that the output state $|\Psi_I^{(out)}\rangle$ is as close as possible to a target state $|\Psi^{(tar)}\rangle$. The fidelity of the state transformation $|\Psi^{(in)}\rangle \to |\Psi^{(tar)}\rangle$ is defined as $f(U) = \left|\langle\Psi_I^{(out)}|\Psi^{(tar)}\rangle\right|^2$ and it occurs with the success probability $s(U) = |\alpha|^2$. The objective, then, is to find a mode transformation $U$ that maximizes success probability while ensuring $f(U)=1$ for the desired logical target state transformation.

A fusion of two linear cluster states $C_n$ and $C_m$ containing qubits $(1,\ldots,n)$ and $(n+1,\ldots,n+m)$ means implementing the state transformation $|C_n\rangle|C_m\rangle \to |C_{n+m}\rangle$. Existing experimental cluster state generation schemes closely follow the original proposal of Raussendorf, Browne, and Briegel [6] where separate photonic qubits or small clusters are fused into a larger cluster state by a sequence of probabilistic optical CZ gates [7]. The CZ gate is conditioned on simultaneous detection of all photons with an overall probability of success 1/9, implying the success probability of $(1/9)^{n-1}$ for fusing $n$ unentangled optical qubits into a linear cluster state. However, in application to linear optical cluster state generation the optimality of the CZ gate as a generating mechanism has not been analyzed previously. Our recent results revealed that this method is far from optimal [8]. Employing numerical optimization based on methods developed in [9], we have identified a method for constructing a linear cluster state $C_{n+1}$ with maximal success probability of $(1/2)^n$ while maintaining unit fidelity. This is in stark contrast to previous analytical results on the optimization of the CZ gate [10] where it was demonstrated that the maximal success rate of the CZ gate is 1/9. Resolving this paradox, we introduce a new theoretical approach for constructing linear optical gates for fusing arbitrary-length clusters.

As our numerical analysis [8] revealed, there exist various multi-mode (i.e. n-qubit) transformations that produce linear cluster states from an initial multi-qubit product state with the same maximal success probability. However, the structure of these solutions is too complex to allow a straightforward decomposition into a chain of two-qubit transformations. We expect that such structure may exist for both the addition of a single qubit and the addition of a Bell pair based primarily on the power-law dependence of the success probability [8].

First, let us consider the simplest case when the first cluster $|C_n\rangle$ contains $n$ qubits and the second cluster $|C_m\rangle$ is just a single qubit i.e., $|C_m\rangle \equiv |+\rangle$, where $|\pm\rangle \equiv (|H\rangle \pm |V\rangle)/\sqrt{2}$ (see Fig.1). In general, the optimization of a linear optical transformation that fuses $|C_n\rangle$ and $|C_1\rangle$ into $|C_{n+1}\rangle$ involves the entire set of $2(n+1)$ modes of $n+1$ qubits. However, the numerical power-law result for the scaling of the success probability, $s=(1/2)^n$ [8] for the $|+\rangle_1 \ldots |+\rangle_{n+1} \to |C_{n+1}\rangle$ transformation ($n=2,\ldots,7$) indicates that a concatenatable transformation acting only on a subset of modes for qubits $n$ and $n+1$ may exist. We denote this operation as $\widetilde{CZ}_1$, where the subscript stands for fusing a single qubit to an existing cluster of arbitrary length. In other words, the fusion of $n+1$ unentangled qubits is a product of $n$ identical two-qubit four-mode $\widetilde{CZ}_1$ operations. Obviously, the success probability of the $\widetilde{CZ}_1$ operation must equal ½, which is greater than the maximum success probability for the optical $CZ$ gate.

It may come as a surprise that the $CZ$ gate and $\widetilde{CZ}_1$ operation perform an identical task with different success probability. An explanation can be found in quantum control theory [11] where one distinguishes two types of control problems. The first type is aimed at constructing a desired transformation (generally referred to as an *operator*) acting on entire Hilbert space. For example, in $\mathbf{C}^4$ the action of a $CZ$ operator is formulated as $|H,H\rangle \to |H,H\rangle$, $|H,V\rangle \to |H,V\rangle$, $|V,H\rangle \to |V,H\rangle$, $|V,V\rangle \to -|V,V\rangle$. The second type (or *state control*) requires designing a quantum transformation affecting only a specific initial state of the system. In contrast, here we consider a hybrid type of transformation acting on a *subspace* of the entire Hilbert space. Note that the $\widetilde{CZ}_1$ operation is of the hybrid kind while the canonical $CZ$ gate is an operator! When "fusing" a *single* qubit with an n-qubit cluster one strictly speaking does not work with operators acting on the two-qubit $\mathbb{C}^4$ space. Since the state of the $(n+1)^{\text{st}}$ qubit is fixed to $|+\rangle$, all transformations, including $CZ$ and $\widetilde{CZ}_1$, are acting only on $\mathbb{C}^2_{in}$, the *subspace* of $\mathbb{C}^4$ spanned by states $|V\rangle_n|+\rangle$ and $|H\rangle_n|+\rangle$. The $\mathbb{C}^2_{in}$ space is being mapped onto two-dimensional subspaces of the full $\mathbb{C}^4$ space. If the action of an operator on $\mathbb{C}^2_{in}$ is identical to the action of a $CZ$ gate on $\mathbb{C}^2_{in}$ then the operator will fuse one qubit to any $C_n$ cluster forming a $C_{n+1}$ cluster state. This operator must satisfy the set of equations determining its action on $\mathbb{C}^2_{in}$:

$$\widetilde{CZ}_1|H\rangle|+\rangle = CZ|H\rangle|+\rangle = |H\rangle|+\rangle$$
$$\widetilde{CZ}_1|V\rangle|+\rangle = CZ|V\rangle|+\rangle = |V\rangle|-\rangle \quad (3a,b)$$

In the context of linear optical entangling gates conditioned on coincidence multimode photon detection, one should further relax the requirement on the $\widetilde{CZ}_1$ operation by adding scaling factors $\alpha$ and $\beta$ to reproduce Eq.(2). To distinguish an abstract $\widetilde{CZ}_1$ operation satisfying equations (3a,b) from a linear optical transformation given by Eq.(2) we use the notation $\widetilde{CZ}_1^{LO}$:

$$\widetilde{CZ}_1^{LO}|H\rangle|+\rangle = \alpha|H\rangle|+\rangle + \beta|\Psi_{II,H}^{(out)}\rangle,$$
$$\widetilde{CZ}_1^{LO}|V\rangle|+\rangle = \alpha|V\rangle|-\rangle + \beta|\Psi_{II,V}^{(out)}\rangle.$$
(4a,b)

To find an optical mode transformation matrix $U$ generating $\widetilde{CZ}_1^{LO}$ Eqs.(4a,b) need to be combined with Eq.(1) resulting in a system of eight polynomial equations in matrix elements $U_{i,j}$. These equations can be solved analytically using the standard Buchberger's algorithm [12], providing the following $4\times 4$ mode transformation matrix $U = A \cdot B \cdot C$ where,

$$A = e^{ix_2\sigma_z^{(1)}} \oplus e^{ix_1\sigma_x^{(2)}} e^{-ix_2\sigma_z^{(2)}}, \quad C = \hat{I}^{(1)} \oplus e^{i\frac{\pi}{4}\sigma_y^{(2)}}$$
$$B_{i,j} = \delta_{(i,j),(4,2)} - \delta_{(i,j),(2,4)} + \delta_{(i,j),(1,1)} + \delta_{(i,j),(3,3)}$$
(5)

and $\hat{I}^{(1,2)}, \sigma_{x,y,z}^{(1,2)}$ are $2\times 2$ matrixes acting on the $H$ and $V$ modes of qubits 1 and 2 respectively, and $\delta$ denotes the Kronecker delta. An essential part of the transformation $U$ is the matrix $B$ which performs the following mode operation: $(a_H^\dagger)_{1,2} \to (a_H^\dagger)_{1,2}$, $(a_V^\dagger)_{1,2} \to \mp(a_V^\dagger)_{2,1}$. Solutions of the form (5), where $x_1$ and $x_2$ are arbitrary real parameters, automatically guarantee fidelity $f(U)=1$ and maximize the success probability $s(U)=1/2$. We remark that the mode transformation is defined in Eq. (1) such that operation $A$ in Eq. (5) precedes operation $B$ and $C$ follows after $B$. Note that the operator $e^{ix_1\sigma_x^{(2)}}$, when acting on the states $|H\rangle|+\rangle$ and $|V\rangle|+\rangle$, adds an overall phase since $\sigma_x|+\rangle = |+\rangle$. The matrix $e^{ix_2\sigma_z^{(1)}} \oplus e^{-ix_2\sigma_z^{(2)}}$ which corresponds to a local two-qubit operator $e^{ix_2(\sigma_z^{(1)}-\sigma_z^{(2)})}$, acts as identity on the space spanned by $|H\rangle|H\rangle$ and $|V\rangle|V\rangle$. Notice that the space spanned by the states $|H\rangle|V\rangle$ and $|V\rangle|H\rangle$ is mapped outside the computational space by the next operation represented by the matrix $B$. Therefore, parameters $x_1$ and $x_2$ do not affect the state transformation.

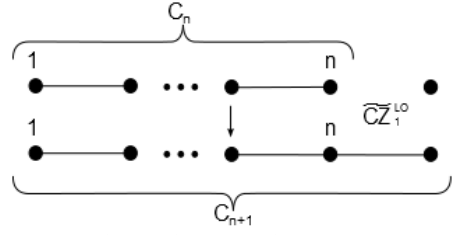

Figure 1. Fusing a single qubit to a $C_n$ cluster.

We would like to emphasize that the sequential character of the overall transformation, such as when the fusion of $n$ qubits is generated by a sequence of four-mode $\widetilde{CZ}_1^{LO}$ operations, is important for cluster state generation. First, a sequential approach reduces the number of physical resources (beam splitters) needed to generate a cluster state. To sequentially generate a linear cluster $|C_n\rangle$ from a product state one only needs $n-1$ polarization-dependent beam splitters. In contrast, a generic $2n$-mode transformation requires a sequence of $n(2n-1)$ beam splitters [13]. Secondly, unlike a global $2n$-mode operation, a sequential transformation can be implemented synchronously with the one-way computation by a sequence of measurements. Therefore, if a measurement fails before the computation is completed, the remaining unused photonic resources can be saved for another attempt. In particular, if the success probability of a single gate equals $s$, the average number of spared photons per successful computation approaches the value of $[n-(2-s)/(1-s)]s^{1-n}$ for large $n$.

Let us now consider the more complex case $m=2$ when the cluster $|C_2\rangle$ is added to $|C_n\rangle$. We again exploit the notion of hybrid operations and generalize the $\widetilde{CZ}_1$ operation to $\widetilde{CZ}_2$. The subscript 2 reflects the fact that two qubits are being added to the cluster. Recall that the state $|C_2\rangle$ is, up to local rotations, equivalent to a Bell state: application of a Hadamard gate to the second qubit in the state $|\Phi^+\rangle$ transforms it into the cluster state $|C_2\rangle \equiv (|\Phi^-\rangle + |\Psi^+\rangle)/\sqrt{2}$.

The action of $\widetilde{CZ}_2$ is analogous to the action of $\widetilde{CZ}_1$ given by Eqs. (3a,b), where the state $|+\rangle$ is now replaced by the state $|C_2\rangle$. Similarly, we can define a linear optical transformation $\widetilde{CZ}_2^{LO}$ as follows,

$$\widetilde{CZ}_2^{LO}|H\rangle|C_2\rangle = \alpha|H\rangle|C_2\rangle + \beta|\Psi_{\perp H}^{(out)}\rangle,$$
$$\widetilde{CZ}_2^{LO}|V\rangle|C_2\rangle = \alpha|V\rangle|C_2^\perp\rangle + \beta|\Psi_{\perp V}^{(out)}\rangle,$$
(6a,b)

where state $|C_2^\perp\rangle = CZ|-\rangle|+\rangle$ is orthogonal to the standard cluster state $|C_2\rangle = CZ|+\rangle|+\rangle$.

Unfortunately, a complete analytical optimization of $\widetilde{CZ}_2^{LO}$ is not possible due to the algebraic complexity of the problem. However, exploiting the idea of hybrid operations we can find an analytical solution, which reproduces our previous result $|\alpha|^2 = 1/4$ obtained by numerical optimization [8]. This solution again provides a higher success probability than $CZ^{LO}$ by factor of $9/4$ [8].

To understand the analytic structure of this optimal solution, we start with an assumption that the fusion of a Bell pair may be partitioned into two concatenated operations as indicated in Fig.2: a polarization beam splitter (PBS) gate [14] acting on qubits $n$ and $n+1$ followed by a "stretch" gate acting on qubits $n+1$ and $n+2$.
The initial $|C_n\rangle$ cluster state has the form $|C_n\rangle = (|C_{n-1}\rangle|H\rangle_n + |\tilde{C}_{n-1}\rangle|V\rangle_n)/\sqrt{2}$, where we use the notation $|\tilde{C}_{n-1}\rangle = \sigma_z^{(n-1)}|C_{n-1}\rangle$. After the action of a PBS transformation on qubits $n$ and $n+1$ (see Fig.2), with success probability $1/2$, the input state for the stretch gate can be cast in the following form,

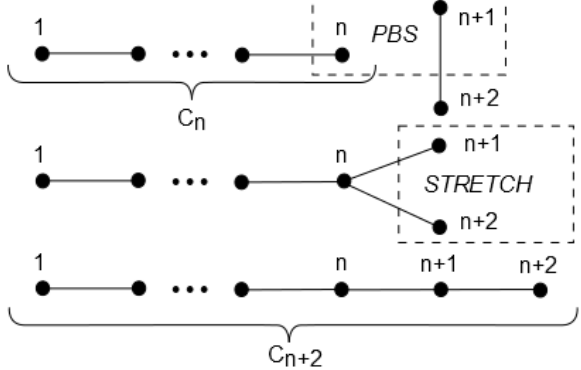

Figure 2. Fusing a Bell pair to a $C_n$ cluster.

$$|\Psi_{in}\rangle = \frac{1}{\sqrt{2}}\left(|C_{n-1}\rangle|H\rangle_n|+,+\rangle_{n+1,n+2} + |\tilde{C}_{n-1}\rangle|V\rangle_n|-,-\rangle_{n+1,n+2}\right).$$

The final, or "target" state of the stretch gate is,
$$|\Psi_{tar}\rangle = \frac{1}{\sqrt{2}}\left(|C_{n-1}\rangle|H\rangle_n|C_2\rangle_{n+1,n+2} + |\tilde{C}_{n-1}\rangle|V\rangle_n\sigma_z^{(n+1)}|C_2\rangle_{n+1,n+2}\right).$$

Interestingly, the stretch gate does not affect the spatial mode $n$; only modes $n+1$ and $n+2$ are involved in the required optical transformations. Therefore, we can consider the action of the stretch gate as a stand-alone hybrid operation on two states:

$$|+\rangle|+\rangle \rightarrow CZ|+\rangle|+\rangle \equiv |C_2\rangle$$
$$|-\rangle|-\rangle \rightarrow CZ|-\rangle|+\rangle \equiv \sigma_z^{(1)}|C_2\rangle$$
(7a,b)

The four-mode transformation matrix for the stretch gate operation can be defined as $U = A \cdot X \cdot B$,

$$X = \begin{pmatrix} \sin(\pi/8) & 0 & -\cos(\pi/8) & 0 \\ 0 & \cos(\pi/8) & 0 & -\sin(\pi/8) \\ \cos(\pi/8) & 0 & \sin(\pi/8) & 0 \\ 0 & \sin(\pi/8) & 0 & \cos(\pi/8) \end{pmatrix},$$

$$A = e^{i\frac{\pi}{4}\sigma_z^{(1)}} \oplus e^{-i\frac{\pi}{4}\sigma_z^{(2)}}e^{-i\frac{\pi}{4}\sigma_y^{(2)}}, \quad B = e^{i\frac{\pi}{4}\sigma_y^{(1)}} \oplus e^{-i\frac{\pi}{4}\sigma_x^{(2)}}.$$

There is also a more general interpretation of the action of the stretch gate where the gate is considered as a tool for transporting entanglement along the cluster (see Fig. 3); the stretch gate implements the operation of moving the central qubit $n$ one position while preserving any form of pre-existing entanglement of the central qubit with an arbitrary external system.

It is also possible to fuse larger linear clusters using the same approach. For instance, for the fusion operations $C_n + C_3 \rightarrow C_{n+3}$ and $C_n + C_4 \rightarrow C_{n+4}$ we numerically found maximal success rates of $1/4$ and $0.153$, respectively. Another example is adding a qubit to the *middle* of a linear cluster state (a "grafting" operation) which normally requires three $CZ$ gates resulting in a success probability of $1/9^3 \approx 0.00137$.

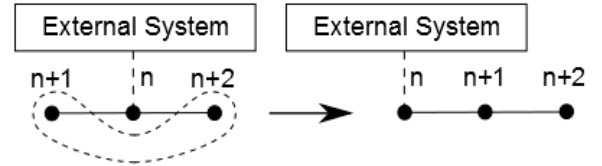

Figure 3. Entanglement swapping by the stretch gate.

When recast in the form of a hybrid operation this gate can be implemented with success probability of $\approx 0.0417$.

One counterintuitive result of our work is that the rate of production of the linear cluster states does not increase when more entanglement resources are invested in the preparation of the initial state. In particular, we have seen that adding a single photonic qubit $|C_1\rangle$ to a cluster has success probability $1/2$, and thus adding two separable photonic qubits to a cluster can be implemented with probability $1/4$, which is the same as the optimal success probability for fusing a Bell pair $|C_2\rangle$ to the same initial cluster.

The idea of hybrid operations developed here has applications beyond linear cluster state generation. For instance, in the case of 2D clusters, a weaving operation [15] can be recast in the form of a hybrid gate.

*Conclusions.* We performed an analytical analysis of the problem of photonic cluster-state generation. We suggested a new scheme that provides the most efficient method of cluster state generation and requires no ancillas. Our analytical results demonstrate that previous methods of cluster state generation are far from optimal. The success probability of our scheme in comparison with traditional schemes grows exponentially with the size of the cluster. We expect that future experiments with photonic clusters will exploit this scheme to provide the most efficient realization of linear optics and quantum information technology.

*Acknowledgments.* Part of this work was performed at ORNL, operated by UT-Battelle for the US DOE under contract no. DE-AC05-00OR22725; DBU acknowledges support from AFRL Information Directorate under grant FA 8750-11-2-0218; DBU and LK acknowledge support from the NSF under grant PHY-1005709; and PMA, MLF, and AMS would like to thank AFOSR for support of this work. Any opinions, findings, and conclusions or recommendations expressed in this material are those of the authors and do not necessarily reflect the views of AFRL.